\documentclass{PoS}

\usepackage{bm}
\usepackage{bbm}
\usepackage{amsmath}
\usepackage{placeins}

\newcommand{\gU}{\mathcal{U}}

\newcommand{\Z}{\mathbb{Z}}

\newcommand{\vev}[1]{\left\langle #1 \right\rangle}

\newcommand{\rep}{\mathcal{R}}

\newcommand{\trP}{\mathcal{P}}

\DeclareMathOperator{\tr}{tr}

\renewcommand{\Re}{\operatorname{Re}}

\newcommand{\ii}{\mathrm{i}}

\newcommand{\includeEPSTEX}[1]{\includegraphics{#1}}

\graphicspath{{publPlots/}}

\title{Phase diagram of the G(2) Higgs model and G(2)-QCD}

\ShortTitle{Phase diagram of the G(2) Higgs model and G(2)-QCD}

\author{\speaker{Bj\"orn H. Wellegehausen}\\
        Theoretisch-Physikalisches Institut, University Jena\\
        E-mail: \email{bjoern.wellegehausen@uni-jena.de}}

\abstract{We investigate gauge theories based on the smallest exceptional simple
lie group $G_2$. Our first model considered here is $G_2$ Yang-Mills coupled to
a fundamental Higgs field. In 4 spacetime dimensions we explore the phase
diagram of the theory, showing that at larger Higgs masses the first order
deconfinement phase transition turns into a crossover and therefore connects
the low temperature confined phase with the high temperature deconfined phase. The second model
investigated is $G_2$ Yang-Mills coupled to fundamental fermions. It shares many
features with QCD, especially fermionic baryons, but due to the absence of the
fermion sign problem we can investigate this theory with Monte Carlo techniques even at low temperature and high baryonic
density. First results on small lattices are presented.}

\FullConference{XXIX International Symposium on Lattice Field Theory \\
		 July 10-16 2011\\
		 Squaw Valley, Lake Tahoe, California}

\begin{document}

\section{Introduction}

\noindent
Due to the fermion sign problem of QCD at real quark chemical potential, Monte
Carlo techniques are not directly applicable to investigate the QCD phase diagram,
especially at low temperatures and high densities \cite{Gattringer:2010zz}.
Understanding the theory in this region of the QCD parameter space is very
important for dense quark systems and the formation of compact stellar objects.
Currently, the only reliable information is obtained from continuum methods,
which usually require approximations or QCD-like theories having as
many features as possible in common with QCD. One example of those theories is two colour two flavour QCD. 
Its phase diagram was explored in a series of papers
\cite{Hands:2000,Hands:2010gd}. The main drawback of two
colour QCD is the absence of fermionic degrees of freedom in the hadron spectrum.
In this work we investigate QCD-like theories based on the exceptional
gauge group $G_2$. It was already shown that $G_2$ Yang Mills theory undergoes a 
first order phase transition from the low temperature confined phase to the high
temperature deconfined phase \cite{Pepe:2006er,Cossu:2008}. Additionally, the
gauge group $SU(3)$ of strong interaction is a subgroup of $G_2$ and this observation has interesting consequences \cite{Holland:2003jy}. With a
Higgs field in the fundamental $7$-dimensional representation one can break the $G_2$ gauge symmetry 
to the $SU(3)$ symmetry via the Higgs mechanism. 
When the Higgs field in the action
\begin{equation}
S[A,\phi]=\int d^4x\left(\frac{1}{4g^2}\tr F_{\mu\nu}F^{\mu\nu} 
+\frac{1}{2}(D_\mu\phi,D_\mu\phi) +V(\phi)\right)\label{SHiggs}
\end{equation}
picks up a vacuum expectation value $v$, 
the $8$ gluons belonging to $SU(3)$ remain massless
and the additional $6$ \emph{gauge bosons} acquire a mass proportional to $v$.
In the limit $v\to\infty$ they are removed from the spectrum such that $G_2$
Yang-Mills-Higgs (YMH) theory reduces to $SU(3)$ Yang-Mills theory. Even more interesting, 
for intermediate and large values of $v$, the $G_2$ YMH-theory mimics $SU(3)$
gauge theory with dynamical 'scalar quarks'. The second theory presented in this
work is $G_2$-QCD \cite{Holland:2003jy} with the action given by
\begin{equation}
S[A,\psi]=\int d^4x\left(\frac{1}{4g^2}\tr F_{\mu\nu}F^{\mu\nu} 
+\frac{1}{2} \bar{\psi}\left(\gamma_\mu\, D^\mu +m+\gamma_0\,\mu\right) \psi
\right)\label{SQCD}.
\end{equation}
Here $\psi$ is a massive Dirac spinor in the fundamental $7$-dimensional
representation of $G_2$. We show that even at finite quark chemical potential
$\mu$ the fermion determinant is non-negative. This allows us to investigate
the phase diagram at zero temperature and finite chemical potential.
Moreover the spectrum contains fermionic baryons and we expect the theory to
behave in many aspects very similar to QCD.

\section{The confinement-deconfinement transition in $G_2$ Yang-Mills}

\begin{figure}
\hskip0.2cm\scalebox{0.86}{\includeEPSTEX{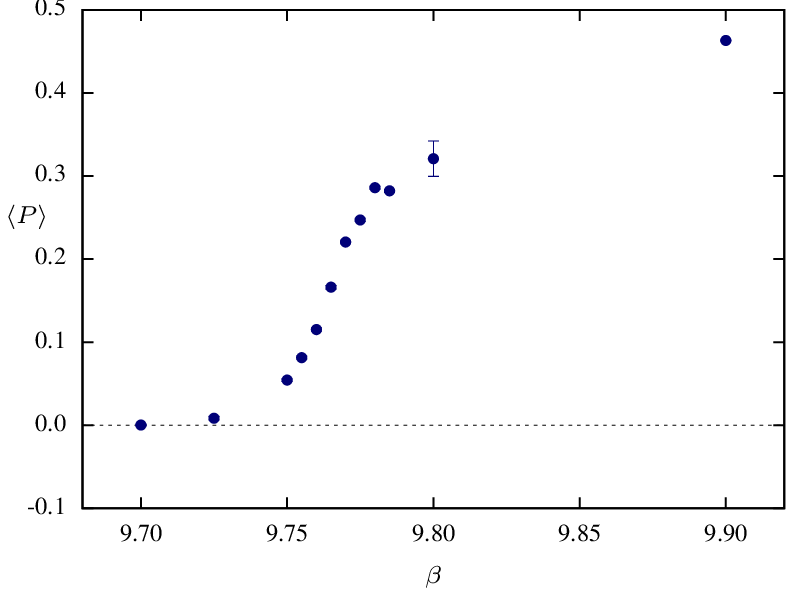}}\hskip 1.0cm
\scalebox{0.86}{\includeEPSTEX{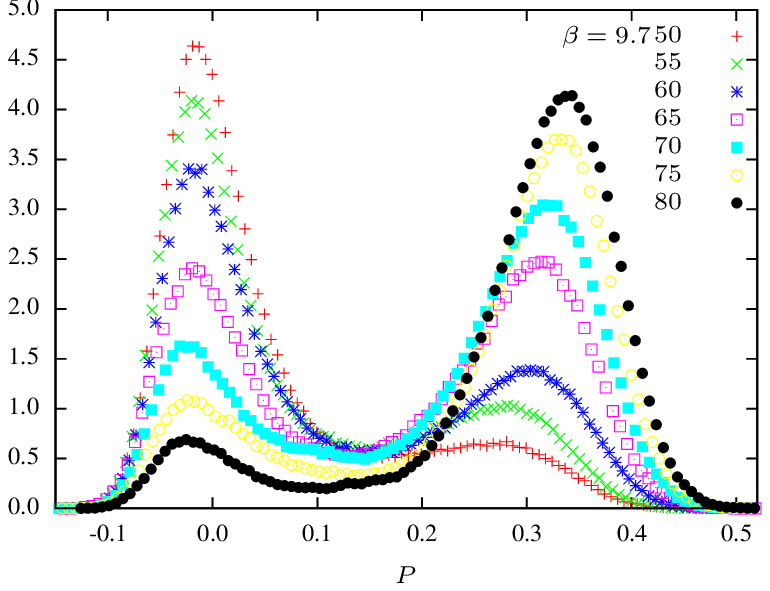}}
\caption{ \textsl{Left panel:} Polyakov loop expectation values at the finite
temperature confinement-deconfinement transition on a $16^3\times 6$ lattice.
\textsl{Right panel:} Histograms of the Polyakov loop in the vicinity of the
phase transition point $\beta_\text{crit}=9.765$, pointing to a first order
deconfinement transition.}
\label{fig:ymTransition}
\end{figure}

\noindent
For a gauge group with non-trivial center the trace
of the Polyakov loop in representation $\rep$
\begin{equation}
P_\rep(\vec{x})=\tr_\rep \trP(\vec{x}), \quad
\trP(\vec{x}) =\frac{1}{N_c}\tr
\left(\exp\; \ii\int_0^{\beta_T}\!\!\!\!
A_0(\tau,\vec{x}) \,d\tau\right),\quad \beta_T=\frac{1}{T},\label{intro1}
\end{equation}
tranforms under \emph{center transformations} like $P_\rep(\vec{x})\mapsto
z^k \, P_\rep(\vec{x})$ where z is an element of the center of the gauge group
and $k$ is the N-ality of the representation $\rep$. In a pure gauge theory the
only dynamical degrees of freedom are gluons in the adjoint center-blind representation. Therefore the Polyakov loop
in any representation with non-zero N-ality is an order parameter for the
spontaneous breaking of center symmetry. On the other hand the vacuum
expectation value of the Polyakov loop is related to the free energy of an
infinitely heavy test quark in the gluonic bath, $\vev{P} \propto e^{-\beta F}$.
Consequently, in the confined phase center symmetry is unbroken, while it is
spontaneously broken in the deconfined phase. If we couple fundamental matter
fields, as for example in QCD, the center symmetry is explicitly broken and the Polyakov loop is not an order parameter
for confinement anymore. But it is still expected that QCD confines
colour in the sense that charges of test quarks are screened by dynamical
light quarks. In $G_2$ gluodynamics the situation is very similar to the QCD
case: Quarks and anti-quarks transform under the fundamental $7$-dimensional
representation. Since the center of $G_2$ is \emph{trivial} these fundamental
charges can be screened by at least three gluons in the adjoint $14$-dimensional
representation,
\begin{equation}
(7) \otimes (14) \otimes (14) \otimes (14)=(1) \oplus \dotsb.
\label{eqn:screening}
\end{equation} 
The Polyakov loop ceases to be an order parameter for confinement. In both, 
$G_2$ gluodynamics and QCD, confinement can be identified with a
linear rising inter-quark potential at intermediate scales and string breaking
at larger distances in every representation of the gauge group
\cite{Greensite:2006sm,Wellegehausen:2010ai}. Nevertheless on a finite lattice
the Polyakov loop serves as an approximate order parameter which changes rapidly at the phase transition and is small (but non-zero) in the
confining phase. We see a clear signal in the Polyakov loop at the
confinement-deconfinement transition and the double peak structure in the
histograms points to a first order confinement-deconfinement transition for
$G_2$ gluodynamics (Fig. \ref{fig:ymTransition}), which is in agreement with
earlier results in \cite{Pepe:2006er,Cossu:2008}.

\section{The $G_2$ Higgs model}

\begin{figure}[htb]
\hskip0.5cm\scalebox{0.85}{\includeEPSTEX{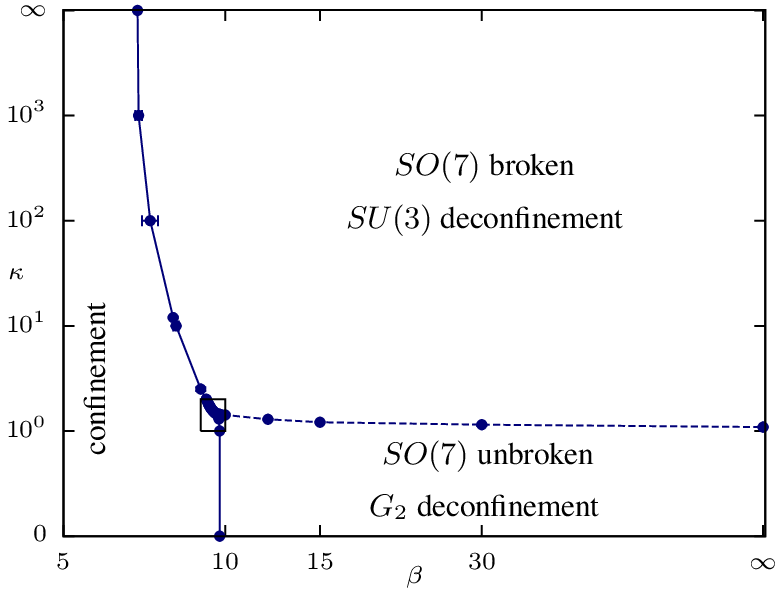}}\hskip1cm
\scalebox{0.85}{\includeEPSTEX{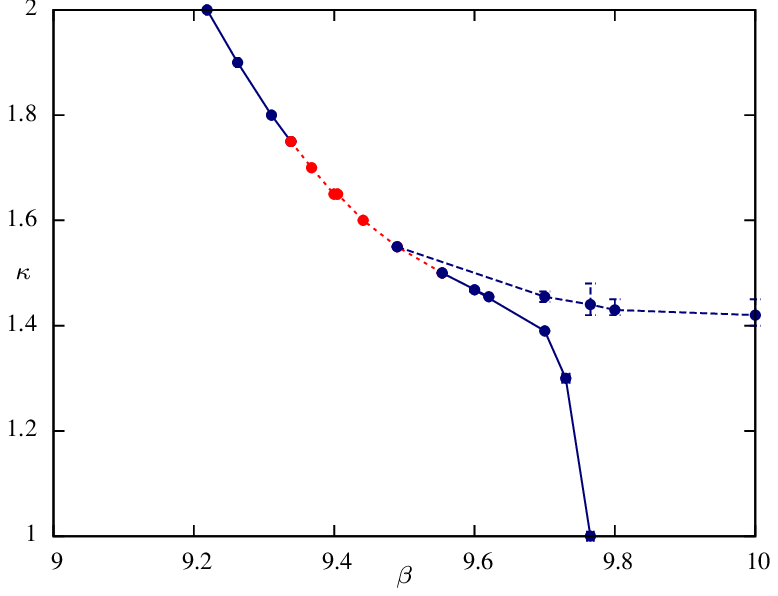}}
\caption{Phase transition lines on a $16^3 \times 6$ lattice.
The solid line corresponds to the first order deconfinement
transitions and the dashed line to the second order Higgs transitions. The plot
on the right panel shows the details inside the small box in the plot
on the left panel where the transition lines almost meet. The dotted line
between the first order lines corresponds to a window where the transition is a
crossover.}
\label{fig:phasediagram16}
\end{figure}

\noindent
The lattice action for the $G_2$ Yang-Mills-Higgs theory
(\ref{SHiggs}) reads
\begin{equation}
S_{\rm YMH}[\,\gU,\Phi] = \beta \sum \limits_\square \left( 1-\frac{1}{7} \tr \Re
\gU_\square \right)-\kappa \sum \limits_{x\mu} \Phi_{x+\hat\mu}\, \gU_{x,\mu}
\Phi_{x},\qquad \Phi_x\cdot\Phi_x=1,\label{latticeaction}
\end{equation}
where $\Phi$ is a seven component normalized real scalar field.
In the limit $\beta\to\infty$ the gauge bosons, belonging to the coset space
$G_2/SU(3) \sim SO(7)/SO(6) \sim S^6$, decouple and the theory reduces to an
$SO(7)$-invariant nonlinear $\sigma$-model. It shows spontaneous symmetry
breaking down to $SO(6)$ at a second order phase transition. With respect to the
$SU(3)$ subgroup of $G_2$ the fundamental representations $(7)$ and $(14)$ branch 
into the following irreducible $SU(3)$-representations:
\begin{equation}
(7) \longrightarrow(3)\oplus (\bar{3}) \oplus (1) \quad,\quad
(14) \longrightarrow(8) \oplus (3) \oplus (\bar{3}).
\end{equation}
\begin{floatingfigure}[htb]
\scalebox{0.35}{\includeEPSTEX{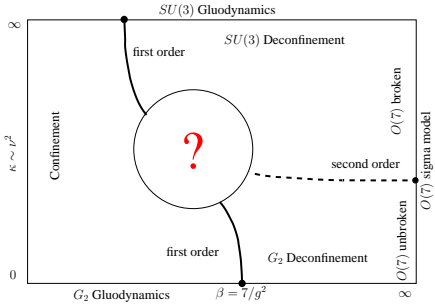}}
\caption{Expected phase diagram in the parameter space $(7/g^2,\kappa)$.}
\label{fig:sketchphases}
\end{floatingfigure}
\noindent
The Higgs field branches into a scalar quark, scalar anti-quark and singlet with 
respect to $SU(3)$. Similarly, a $G_2$-gluon branches into a massless $SU(3)$-gluon
and additional gauge bosons with respect to $SU(3)$. The latter eat up the
non-singlet scalar fields such that the spectrum in the broken phase
consists of $8$ massless gluons, $6$ massive gauge bosons and one
massive Higgs particle. In the limit of $\kappa \to \infty$ these $6$ massive gauge bosons become
infinitely heavy and the theory reduces to $SU(3)$ Yang Mills theory. If
$\kappa$ is lowered, in addition to the $8$ gluons of $SU(3)$, the $6$ additional gauge bosons of $G_2$ begin to participate in the dynamics. 
Similarly as dynamical quarks and anit-quarks in QCD, they transform in the
representations $(3)$ and $(\bar 3)$ of $SU(3)$ and thus explicitly break the $\Z_3$ center
symmetry. As in QCD they are expected to weaken the deconfinement phase
transition \cite{Pepe:2006er}. For $\kappa=0$ we recover $G_2$ gluodynamics with
a first order deconfinement phase transition. Fig.~\ref{fig:sketchphases} shows
the expected phase diagram. Our results for the complete phase diagram in the $(\beta,\kappa)$ 
plane as calculated  on $16^3\times 6$ up to $24^3\times 6$ lattices are summarized in 
Fig.~\ref{fig:phasediagram16}. We calculated histograms and susceptibilities of
the Polyakov loop, the plaquette density and the Higgs action density near the
marked points on the transition lines in this figure. From earlier calculations
on smaller lattices we suggest that a triple point exists. An extrapolation to
the point where the confined phase meets both deconfined phases leads to the
couplings $\beta_\text{trip}=9.62(1)$ and $\kappa_\text{trip}=1.455(5)$. Near this  point the deconfinement transition is very weak, 
continuous or absent. Therefore we performed high-statistics simulations to
investigate this region in parameter space more carefully. Up to a rather small region
surrounding $(\beta_\text{trip},\kappa_\text{trip})$ we can
show that the deconfinement transition is first order and
the Higgs transition is second order. But in this small window in parameter
space, around the \emph{would-be} triple point, the deconfinement transition is
either second order or absent \cite{Wellegehausen:2011ai}.

\section{$G_2$-QCD}
\begin{figure}[htb]
\hskip0.5cm\scalebox{0.85}{\includeEPSTEX{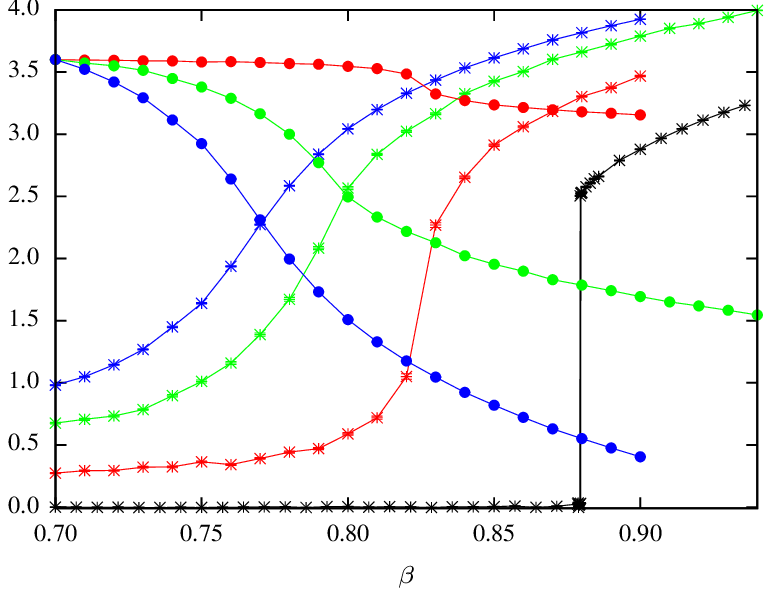}}\hskip1cm
\scalebox{0.80}{\includeEPSTEX{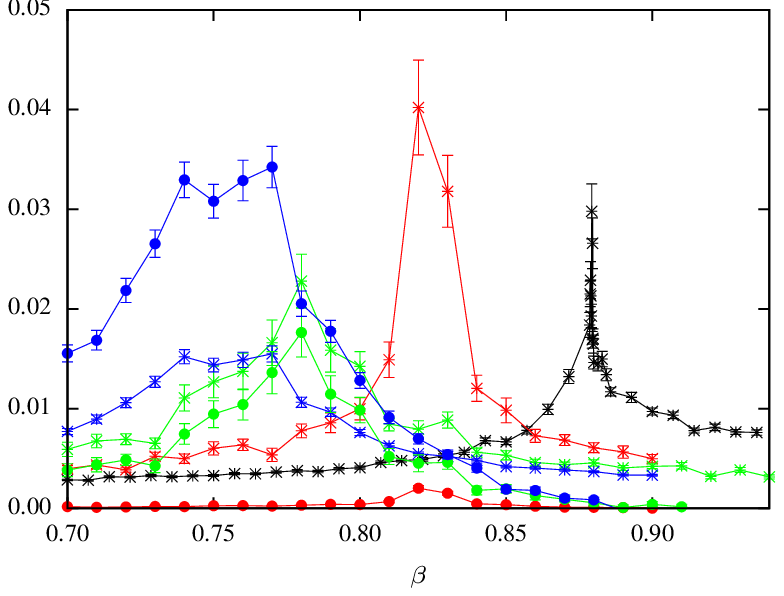}}
\caption{\textsl{Left panel:} Polyakov loop (stars) and (rescaled)
chiral condensate (dots) at finite temperature
and zero chemical potential for $\kappa=0,\,0.096,\,
0.131,\, 0.147$. \textsl{Right panel:}
Susceptibility of the Polyakov loop and chiral
condensate.}\label{fig:finiteTemperatureQCD}
\end{figure}

\begin{figure}[htb]
\scalebox{0.63}{\includeEPSTEX{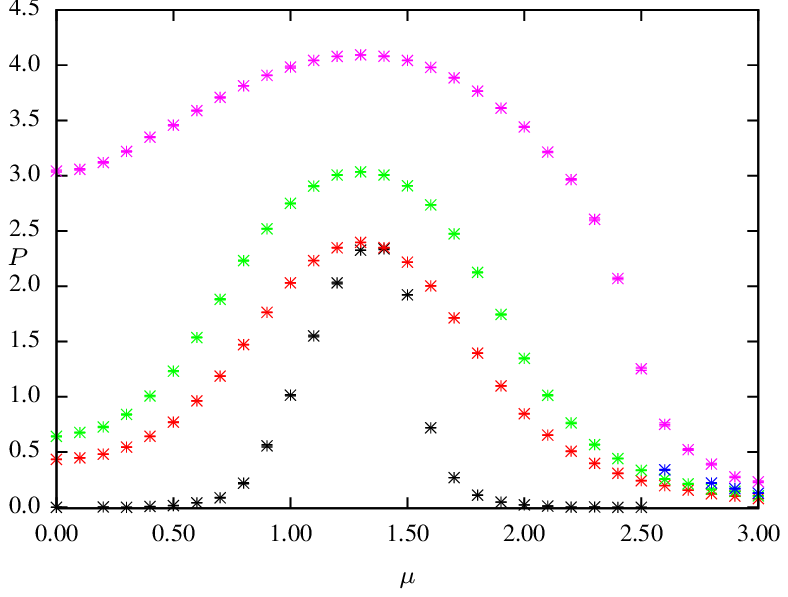}}
\scalebox{0.63}{\includeEPSTEX{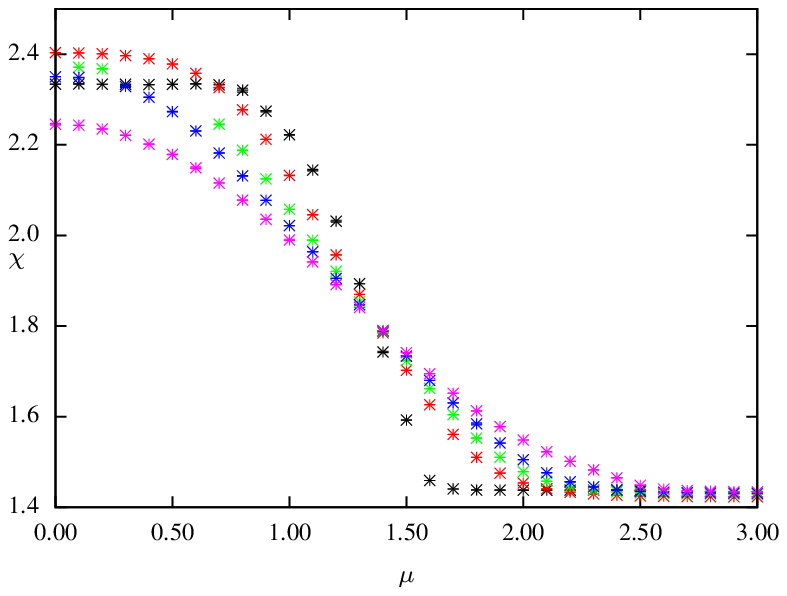}}
\scalebox{0.63}{\includeEPSTEX{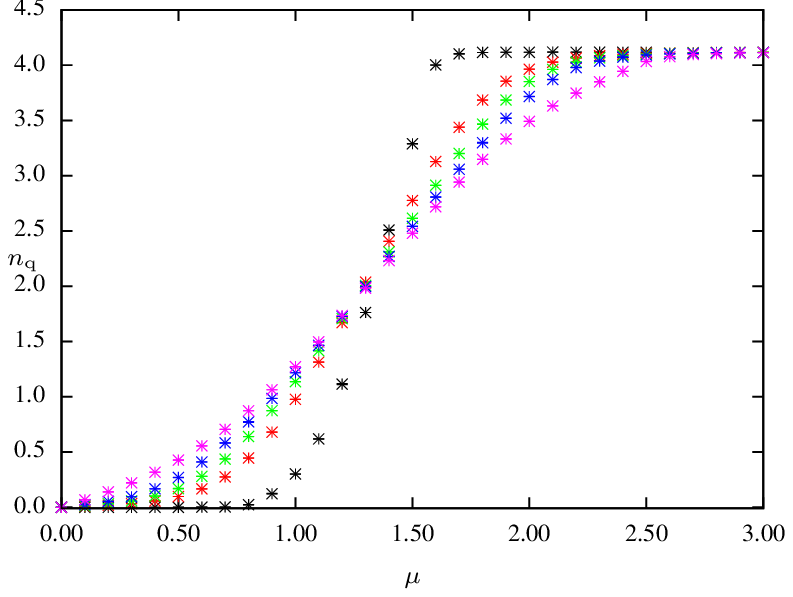}}
\caption{Polyakov loop $P$ (left panel), chiral
condensate $\chi$ (center panel) and quark number
density $n_{\rm q}$ (right panel) for different
temperatures.}\label{fig:differentTQCD}
\end{figure}

\noindent
The lattice action for $G_2$-QCD
(\ref{SQCD}) reads
\begin{equation}
S_{\rm QCD}[\,\gU,\Psi] = \beta \left(S_\text{Sym}[\,\gU]-\frac{1}{2} \sum
\limits_{xy} \bar{\Psi}_{x}\,D_{xy}[\,\gU,m,\mu]\Psi_{y}
\right),\label{latticeaction1}
\end{equation}
where $S_\text{Sym}$ is the Symanzik improved gauge action and
$D[\,\gU,m,\mu]$ the Wilson-Dirac operator at bare fermion mass $m$ and
real quark chemical potential $\mu$.
If a unitary operator $T$ exists such that the Dirac operator obeys the relation
$D^* \,T=T\,D$ with $T^*\,T=-\mathbbm{1}$ and $T^\dagger=T^{-1}$, the fermion
determinant is non-negative \cite{Hands:2000}. For Wilson
fermions this implies the conditions $T \gU T^\dagger=\gU^*$ and $T
\gamma_{\mu}T^\dagger=\gamma_{\mu}^*$. Since every representation of $G_2$ is real, we immediately find (for an
Euclidean representation of the $\gamma$-matrices and charge conjugation matrix
$C$)
\begin{equation}
T=\mathbbm{1}\otimes C\gamma_5 \quad \Rightarrow
\quad T^*\,T=-\mathbbm{1} \quad \Rightarrow \quad \det D[\,\gU,m,\mu]\geq 0.
\end{equation}
This feature of the fermion determinant makes Monte Carlo techniques applicable
and allows us to investigate the complete phase diagram of
$G_2$-QCD on the lattice. The chiral symmetry of $N_f$ flavor QCD is $SU(N_f)_L \otimes SU(N_f)_R \otimes U(1)_B$. 
Since the quark representation for $G_2$ is real the dirac
spinor decomposes into two Majorana spinors for vanishing mass. Consequently, we
find an additional \emph{flavour} symmetry and the continuous $U(1)_B$ turns
into the discrete group $\mathbbm{Z}(2)$. The chiral symmetry of $G_2$-QCD is
then \cite{Holland:2003jy}
\begin{equation}
SU(2N_f)_{L=R^*} \otimes \mathbbm{Z}(2)_B.
\end{equation}
Due to the discrete baryon number symmetry we can only distinguish between
states with an even or odd number of quarks. Similar to QCD, mesons
consist of an even number of quarks. Baryons are fermionic bound states with an
odd number of quarks, for example, three quarks or a single quark and three gluons. Our first simulations are performed on rather small lattices with a
spatial extend of $L=8$ and $T=2$ for finite temperature and on a $8^4$ lattice in the case of
zero temperature. The observables considered in this preliminary study are the
Polyakov loop $P$, the chiral condensate $\chi=\frac{1}{V}\frac{\partial \ln
Z}{\partial m}$ and the quark number density $n_{\rm
q}=\frac{1}{V}\frac{\partial \ln Z}{\partial \mu}$. 
First we compare the finite temperature phase transition at zero chemical potential to the phase transition in pure gauge theory. 
Our results for the Polyakov loop and the chiral condensate are shown in
Fig.~\ref{fig:finiteTemperatureQCD} (left panel). 
If we increase the hopping parameter from $\kappa=0$ to $\kappa=0.147$, the
transition in the Polyakov loop becomes weaker and the critical temperature decreases.
This behaviour is also seen in the chiral condensate. The difference between its
value in the chirally symmetric and broken phase grows with decreasing mass,
due to the smaller explicit breaking of chiral symmetry. In Fig.~\ref{fig:finiteTemperatureQCD}
(right panel) we show the
\begin{floatingfigure}[htb]
\scalebox{0.75}{\includeEPSTEX{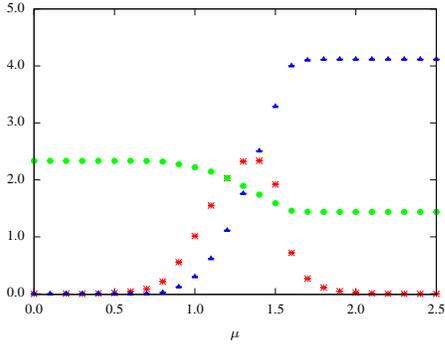}}
\caption{Polyakov loop (stars), chiral
condensate (dots) and quark number density
(triangles) at zero temperature.}\label{fig:zeroTemperatureQCD}
\end{floatingfigure}
\noindent
susceptibilites of both quantities.
The peak becomes very broad for smaller quark masses, pointing to a cross\-over.
On these rather small lattices and within the given resolution in temperature, we can see no difference in the position of the peak. But whether the
deconfinement and chiral transition critical temperature coincide and whether
the transition really is a crossover at smaller quark masses has to be
investigated in further simulations on larger lattices. 
At zero temperature the system stays in the vacuum until at approximately
$\mu=0.8$ the quark number density increases, Fig.~\ref{fig:zeroTemperatureQCD}. 
At $\mu \approx 1.5$ the system is saturated in the sence that each lattice site is occupied by an odd number of
quarks. We also observe a transition in the Polyakov loop together with a drop
in the chiral condensate. This points to a phase transition, or a crossover,
from a confined phase to a chirally symmetric deconfined phase. At larger values
of the chemical potential the Polyakov loop decreases again. This may be an
lattice artifact due to the saturation of the system and has to be investigated
further. If we increase the temperature again, the transition becomes weaker as shown in Fig.~\ref{fig:differentTQCD}. 

\section{Conclusions}

\noindent
We have investigated QCD-like theories based on the gauge group $G_2$. If we
couple $G_2$ Yang-Mills to a fundamental Higgs field, the theory mimics, in the
broken phase, QCD with \emph{scalar} \emph{quarks} and \emph{anti-quarks} with respect to the $SU(3)$ subgroup of $G_2$. In the limit
of infinitely heavy \emph{quarks} we recover $SU(3)$ gauge theory with a first
order deconfinement phase transition. If we decrease the mass the transition
becomes weaker and, in a very small window in parameter space, turns into a crossover.
This behaviour is already very similar to QCD. The second theory investigated is
one flavour $G_2$-QCD which has many features in common with QCD. In
this first study we show that the finite temperature phase transition becomes
weaker with decreasing mass of the quarks. At zero temperature and finite
chemical potential we find a transition in the Polyakov loop and the chiral
condensate together with an increasing baryon number density. First results on
larger lattices will be published in a forthcoming paper
\cite{Wellegehausen:2012}.

\begin{acknowledgments}
\noindent
The author thanks Andreas Wipf, Axel Maas and Christian Wozar for collaboration
and support. Helpful discussions with Lorenz von Smekal, Christof Gattringer,
Kurt Langfeld, Uwe-Jens Wiese and \v{S}tefan Olejn\'{i}k are gratefully acknowledged. This
work has been supported by the DFG under GRK~1523. The simulations in this paper were carried out at the Omega-Cluster 
of the TPI.
\end{acknowledgments}

\end{document}